\documentclass[aps,prd,preprint,floatfix,superscriptaddress]{revtex4-2}
\usepackage{graphicx} % Required for inserting images
\usepackage{amssymb}
\usepackage{amsmath}
\usepackage{bbm}
\usepackage{xcolor}
\usepackage{multirow}                         
\usepackage{slashed}
\usepackage{braket}
\usepackage{mathtools}
\usepackage{subcaption}
\usepackage{cancel}
\usepackage{tikz}
\usepackage{float}
\usepackage{bm}
\usepackage[compat=1.1.0]{tikz-feynman}

\begin{document}

\title{A Density Functional for Quark Matter with Screened Confinement Interactions}
\author{Oliver Heymer}
\affiliation{Institute of Theoretical Physics, Technical University Bergakademie Freiberg, Germany}
\affiliation{Institute of Theoretical Physics, University of Wroclaw, Poland}
\author{David Blaschke}
\affiliation{Institute of Theoretical Physics, University of Wroclaw, Poland}
\affiliation{Center for Advanced Systems Understanding (CASUS) G\"orlitz, HZDR, Germany}
%\author{Oleksii Ivanytskyi}
%\affiliation{Incubator of Scientific Excellence, 
%Center for Simulations of Superdense Fluids,
%University of Wroclaw, Poland}
%
\date{\today}

\begin{abstract}
We derive a density functional for quark matter with linear confinement interaction, applying the principle of saturation of color interactions. For the saturation parameter $\mu$ the ansatz $\mu=k_F$ is suggested, where for the Fermi momentum $k_F \propto n^{1/3}$ going to zero the Cornell potential is obtained and at finite quark densities $n$ an interaction energy density functional $\propto n^{2/3}$ results. 
\end{abstract}
\maketitle

\section{Introduction}

One of the main challenges in the description of strongly interacting matter under extreme conditions is a microscopic description of quark matter in the vicinity of the transition to its hadronic (clustered) phase, where the binding of quarks into their hadronic clusters is governed by confining interactions.
The description of hadrons (mesons, baryons, exotica) in free space as isolated two-, three-, or multiquarks bound states has been advanced in relativistic quark potential models where the pairwise interaction of quarks is given by a confining interaction potential first suggested by the Cornell group \cite{Eichten:1974af,Eichten:1978tg}
\begin{equation}
\label{eq:cornell}
    V(r)=-\frac{\alpha_{\rm eff}}{r} + \sigma r + V_0
\end{equation}
between massive constituent quarks, where $\alpha_{\rm eff}$
is the effective fine structure constant of the strong interaction in a color neutral quark-antiquark state, $\sigma$ is the string tension and $V_0$ is a possible shift of the continuum edge.

%\color{red}What is $\alpha_{\mathrm{eff}}$, $\sigma$, $V_0$?\color{black}

This model was an improvement over the bag model 
\cite{Chodos:1974je,Chodos:1974pn}
of quark confinement since it could well describe the details of the mass spectrum of  hadrons and their electromagnetic decays, in particular in the heavy quark sector. Further details can be taken from Ref. \cite{Lucha:1991vn}. 

For the description of a quark plasma as an interacting many-body system, however, the unscreened Cornell potential 
\eqref{eq:cornell} as such is not suitable due to its divergence at small and large distances. 
The application of the Mayer cluster expansion to determine the coefficients of the (quantum) virial expansion, for instance, would require a potential that saturates for large distances at a constant value so that its continuum edge could be renormalized to zero by subtraction.
Such a behaviour of the interaction potential between color charges is physically justified by the effect of string breaking which occurs when the potential energy stored in the string exceeds twice the rest-mass energy of the lightest constituent quarks so that a light quark-antiquark pair would be created instead of stretching the string further. 
The linear confinement part of the interaction potential would then be replaced by a saturated interaction, e.g.,
%
%\color{red}Here cite Shuckla et al.\color{black}
%
\begin{eqnarray}
\label{eq:conf-sat}
    V_{\rm conf}({r}) &=& \lim_{r\to\infty} \frac{\sigma}{\mu} \left(1- e^{-\mu r }\right) = \frac{\sigma}{\mu} = \mathrm{const},
\end{eqnarray}
%\color{red}
%Should be $\mu\to\infty$, right? Introduce $\bm{r}$, $r$, $\sigma$, $\mu$
%\color{black}
%
which for small distances still resembles the linear confinement, see Fig. \ref{fig:V-r}.
Such a form of the potential is well motivated by the heavy-quark potential obtained in Monte-Carlo simulations of the free energy of lattice QCD with static color charges placed at varying distances \cite{Doring:2005ih}. 
One observes that the constant value where the potential given in equation \eqref{eq:conf-sat} saturates depends on the temperature and chemical potential. This is expected because of the chiral symmetry restoration which lowers the quark masses and also because of the modification of the nonperturbative QCD vacuum effects which would entail a medium dependent lowering of the string tension $\sigma$. 

In the main body of this work, we want to derive a density functional for the saturated confinement interaction that will be applicable for the description of srongly interacting quark matter in the vicinity of the hadron-to-quark matter transition in low-energy QCD.

However, before turning to this task, one has to mention another concept for arriving at a thermodynamic description of quark matter with confining interactions.
The concept of desire is the principle of saturation of color confinement interactions within the range of nearest neighbors.
The nearest neighbor argument has been used by Olive \cite{Olive:1982we} to suggest a density functional for the interaction energy in quark matter at the quark density $n_q$ by simply replacing $r \to n_q^{-1/3}$ in the confinement interaction \eqref{eq:cornell}.
This argument has been formalized subsequently in a more quantitative way by multiplying the bare confinement potential with the nearest-neighbor distribution function which in homogeneous quark matter for $N_c$ colors is simply given by
\cite{Blaschke:1984yj,Ropke:1986qs}
\begin{equation}
    c(r) = \exp{(-4\pi n_q r^3/(3N_c))}.
\end{equation}

A numerical simulation of the quark matter equation of state with the color saturation principle has been pioneered in Ref. \cite{Horowitz:1985tx}, where the name "string-flip model" has been coined.
It characterizes the model assumption that for a given configuration of quark positions the configuration of strings between them is established  instantaneously (by string-flip processes) so that the system is color neutral and its potential energy attains a minimum.
In Ref. \cite{Ropke:1986qs} it has been suggested that the string-flip model could be approximately realized by a system of quasiparticle quarks which obey a Fermi distribution function with a quasifree dispersion relation that is subject to a density-dependent quasiparticle energy shift in Hartree approximation \cite{Blaschke:1984yj,Barter:1992rg}
$\Delta_H
%^{\rm Hartree}
=n_q\int d^3r V(r)c(r)$.
%\color{red}
%The latter sentence is a bit too long for my taste. Furthermore, I'm not sure if the equation is really necessary.
%\color{black}
This quantum statistical description of quark matter has been advanced subsequently also to describe results of lattice QCD thermodynamics by, e.g., Yukalov et al. \cite{Yukalov:1997jk} and Khvorostukhin et al. \cite{Khvorostukin:2006aw}.

For a relativistic description, the energy shift of the scalar confinement potential has been reinterpreted as a density-dependent quark mass and in this form defines a large class of density-dependent quark mass (DDQM) models, e.g.
\cite{Benvenuto:1989kc,Lugones:1995vg,Peng:1997cs,Peng:1999gh,Peng:2000ff,Wen:2005uf,Xia:2014zaa,Li:2015ida}.
%,Issifu:2024lgj}.
%\color{red} Citation error has to be fixed! \color{black}

A systematic approach to such models has been provided within the Path-integral approach to the thermodynamic potential
by Kaltenborn et al. \cite{Kaltenborn:2017hus},
which in Ref. \cite{Ivanytskyi:2022oxv} has been formulated in a chiral-invariant way and extended by implementing vector interaction and color superconducting channels.

\iffalse
========

- density-dependent quark mass model \cite{Wen:2005uf,Li:2015ida,Peng:1997cs,Peng:2000ff,Peng:1999gh,Xia:2014zaa}

- string-flip model \cite{Blaschke:1984yj,Ropke:1986qs,Voss:1988hh,Barter:1989cq,Barter:1992rg}

- relativistic version \cite{Kaltenborn:2017hus,Ivanytskyi:2022oxv}

- calculation in momentum space, ill-posed problem \cite{Shukla:2025egs}

- regularized confinement potential model and gap equation \cite{LeYaouanc:1983huv, Karsch:1987pv}

- screened heavy quark confining potential from lattice QCD  \cite{Doring:2005ih}

- nearest neighbor argument for density-dependent 
\cite{Olive:1982we}

- screening function for effective density dependent selfenergy \cite{Yukalov:1997jk}

\color{red}
\section{TO DO}
\begin{itemize}
    \item Emphasize that n(r) is just the density of one flavor sector
    \item Mention Buckingham potential
    \item Lattice QCD heavy quark potential    
\end{itemize}
\color{black}
\fi

The key element for all these confining density functional approaches is the justification of an energy density that is proportional to density to the power of two-thirds. The latter is provided in this work. 

\section{Theory}

\subsection{The Thomas-Fermi Model for homogeneous particle systems}
Perhaps the oldest and simplest Density functional theory is the Thomas-Fermi model \cite{Thomas1927, Fermi1928}. Here, the total energy is an explicit functional of the particle density $n(\bm{r})$ \cite{Martin2020}. Following Ref. \cite{Martin2020}, it is decomposed into a kinetic energy functional $T[n]$, a Hartree contribution $E_\mathrm{H}[n]$ that describes the interactions among the particles of interest and an external contribution $E_\mathrm{ext}[n]$ that captures the interactions with the rest of nature.

As we consider an isolated system of particles, the external contribution can be neglected. The aim of this work is a description of the interaction between particles. For this reason we will focus solely on the Hartree contribution
\begin{equation}
    E_\mathrm{H}[n] = \int d^3\bm{r}\, V_\mathrm{H}[n](\bm{r}) n(\bm{r})
\end{equation}
which contains the Hartree electrostatic potential
\begin{equation}
    V_\mathrm{H}[n](\bm{r}) = \frac{1}{2}\int d^3\bm{r}'\, W(\bm{r} - \bm{r}') n(\bm{r}')
\end{equation}
For now, the translation invariant interaction kernel $W(\bm{r} - \bm{r}')$ is left unspecified and is going to be addressed in the next section.

We define the difference vector between two particles as $\bm{R}:= (\bm{r} - \bm{r}')$ and in turn obtain the following expression for the Hartree energy functional:
\begin{equation}
    E_\mathrm{H}[n] = \frac{1}{2} \int d^3\bm{R} \int d^3\bm{r}'\; n(\bm{R} - \bm{r}')W(\bm{R}) n(\bm{r}' - \bm{R})
\end{equation}
For homogeneous systems, it is irrelevant at which position the densities are evaluated. They are translation invariant as well. This means that any shift by $\bm{r}'$ is irrelevant and can be set to zero:
\begin{equation}
    E_\mathrm{H}[n] = \frac{1}{2} \int d^3\bm{R} \int d^3\bm{r}'\; n(\bm{R})W(\bm{R}) n(- \bm{R})
\end{equation}
The integral over $\bm{r}'$ evaluates to the system's volume $\Omega$ which turns the Hartree contribution into the following form:
\begin{equation}
    E_\mathrm{H}[n] = \frac{\Omega}{2}\int d^3\bm{R}\, n(-\bm{R}) W(\bm{R}) n(\bm{R})\label{equ:hartree-energy}
\end{equation}
One may go to momentum space and perform a mode expansion for the local densities and for the interaction kernel respectively:
\begin{equation}
    n(\bm{r}) = \int \frac{d^3\bm{k}}{(2\pi)^3} \; n(\bm{k}) e^{-i \bm{k}\cdot \bm{r}} \qquad W(\bm{r}) = \int \frac{d^3\bm{k}}{(2\pi)^3} \; W(\bm{k}) e^{-i \bm{k}\cdot \bm{r}}
\end{equation}
In turn the Hartree contribution writes as follows:
\begin{equation}
    E_\mathrm{H} = \frac{\Omega}{2}\int \frac{d^3\bm{k}}{(2\pi)^3} \int \frac{d^3\bm{q}}{(2\pi)^3}\, n(\bm{k}) W(\bm{k}-\bm{q}) n(\bm{q})
\end{equation}
It should be noted that the densities $n(\bm{k})$ measure the occupancy at a given point $\bm{k}$ in momentum space.
Using the homogeneity argument again, they can be identified it with the Fermi-Dirac distribution $f[\varepsilon](\bm{k})$ which for now is a functional of the single-particle energy dispersion $\varepsilon(\bm{k})$:
\begin{equation}
    f[\varepsilon](\bm{k}) = \frac{1}{1+e^{\beta \varepsilon(\bm{k})}}
\end{equation}

In order to show the implications the homogeneity has, one has to expand $n(\bm{k})$ in the position space basis
\begin{equation}
    n(\bm{k}) = \int d^3\bm{r}\; n(\bm{r}) e^{i\bm{k}\cdot \bm{r}}
\end{equation}
and assume that the local density is equal to the position independent expectation value of itself $\braket{n(\bm{r})}$. Because the latter is a momentum space integral over the Fermi-Dirac distribution we get that
\begin{equation}
    n(\bm{k}) = \braket{n(\bm{r})} \delta^{(3)}(\bm{k}) = \delta^{(3)}(\bm{k})\int \frac{d^3\bm{q}}{(2\pi)^3} \; f[n](\bm{q})
\end{equation}
which leads directly to the following expression:
\begin{equation}
    E_\mathrm{H} = \frac{\Omega}{2}\int \frac{d^3\bm{k}}{(2\pi)^3} \int \frac{d^3\bm{q}}{(2\pi)^3}\, f(\bm{k}) W(\bm{k}-\bm{q}) f(\bm{q})
\end{equation}
As a consequence of requiring a homogeneous density, the Hartree energy is no longer a density functional.
This is a great point to stop. In the thermodynamic limit of an infinite volume, the Hartree contribution diverges. Therefore, it is sensible to divide both sides by $\Omega$ and to work with the Hartree energy density $\epsilon_{\mathrm{H}}$ instead:
\begin{equation}
    \epsilon_\mathrm{H} = \frac{1}{2}\int \frac{d^3\bm{k}}{(2\pi)^3} \int \frac{d^3\bm{q}}{(2\pi)^3}\, f(\bm{k}) W(\bm{k}-\bm{q}) f(\bm{q})\label{equ:hartree-energy-dens}
\end{equation}

\subsection{Regularization of the linear confinement interaction kernel}
We are now in the position where we can wonder about the interaction kernel.
For this, consider the linear confinement kernel which is given by the following equation:
\begin{equation}
    W(\bm{r}) = \sigma r
\end{equation}
In this case $\sigma$ is the string tension and $r$ is the absolute value of the position vector.
What is the problem with this interaction kernel? Similarly to the Coulomb potential, the three dimensional Fourier transform diverges. For this reason, we regularize the kernel following Ref. \cite{Karsch:1987pv}:
\begin{equation}
\label{eq:cornell-sc}
    W(\bm{r})= \lim_{\mu\to 0}\frac{\sigma}{\mu}\left(1- e^{-\mu {r}}\right)
\end{equation}
By induction one can show that the following regularization schemes regularize the kernel as well:
\begin{align}
    W_n(\bm{r}) := \lim_{\mu\to 0}\frac{\sigma}{\mu^n} \frac{1}{r^{n-1} }\left(1- e^{-\mu^n r^n }\right)
\end{align}
However, there are far more regularization schemes that have different functional forms. As long as one takes $\mu\to 0$ limit after performing the Fourier transformation, all regularization schemes will lead to the same result - a diverging Fourier transform of the linear interaction kernel. This is unsatisfying. Instead of throwing in the towel, one could give some physical meaning to the regularization parameter $\mu$ and drop the limit.

Following Ref. \cite{Ropke:1988zza} one may interpret the regularization parameter as a Debye screening parameter. This implies that it is inversely proportional to some Debye radius $r_\mathrm{D}$.

It should be noted that from now on all calculations are carried out with explicit focus on the $n=1$ regularization scheme.

The kernel then turns into
\begin{align}
    W(\bm{r}) := \frac{\sigma}{\mu} \left(1- e^{-\mu r }\right)
\end{align}
which is motivated by the aforementioned lattice QCD calculations \cite{Doring:2005ih}.

\begin{figure}[htb]
%    \centering
    \includegraphics[width=\linewidth]{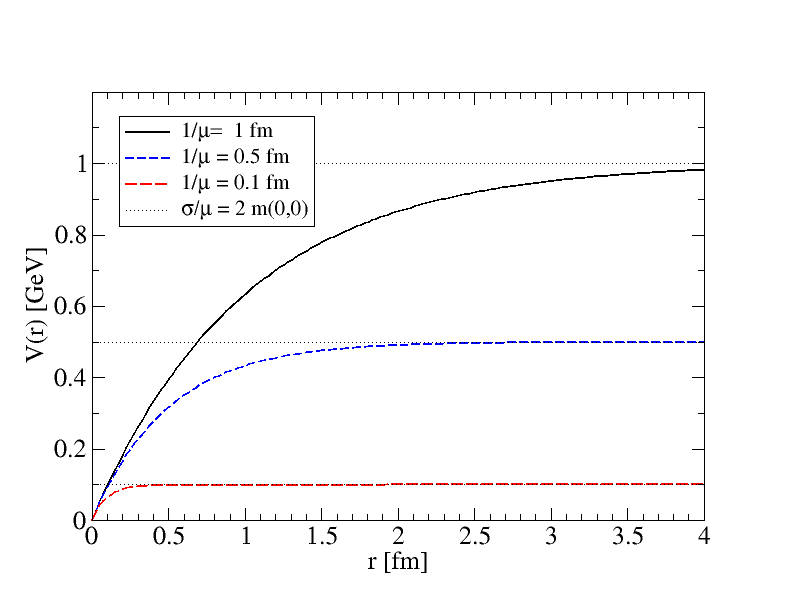}
    \caption{Regularized linear confinement potential for different values of the Debye screening length $r_D=1/\mu$ which introduces a continuum edge $V_{\rm cont}=\sigma/\mu$ (dotted line). }
    \label{fig:V-r}
\end{figure}

In Fig.~\ref{fig:V-r}, the potential is shown as a function of the inter-quark distance for different values of the regularization parameter $r_D=1/\mu$.

\subsection{Fourier transform of the regularized linear confinement potential}
\begin{figure}[htb]
%    \centering
    \includegraphics[width=\linewidth]{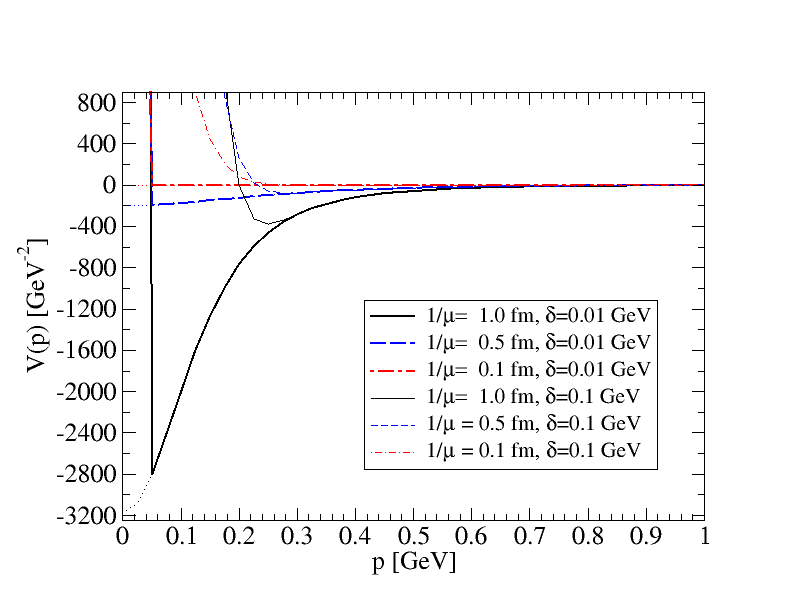}
    \caption{Fourier transform of the regularized linear confinement potential for different values of the Debye screening length $r_D=1/\mu$.
    For sake of illustration, the delta function at the origin is represented by a normalized Gaussian with a width $\delta$.}
    \label{fig:V-p}
\end{figure}

The Fourier transform of this kernel (\ref{eq:cornell-sc}) is (see appendix \ref{app:fourier})
\begin{equation}
    W(\bm{k}) = \frac{\sigma}{\mu} \left[(2\pi)^3\delta(\bm{k})^{(3)} - \frac{8\pi \mu}{(\mu^2 + k^2)^2}\right]\label{equ:fourier-kernel}
\end{equation}
It is shown in Fig. \ref{fig:V-p} for the same values of the screening parameter $\mu$ that were used in Fig. \ref{fig:V-r}. The delta function at the origin is represented by a normalized Gaussian with a width $\delta$.

\subsection{Closed form Hartree Energy Density at Absolute Zero}
We would like to work at absolute zero in which case the Fermi-Dirac distributions in equations \eqref{equ:hartree-energy} and \eqref{equ:hartree-energy-dens} turn into Heavyside step functions $\Theta(k_F-\bm{k})$ and $\Theta(q_F-\bm{q})$.
By inserting \eqref{equ:fourier-kernel} into \eqref{equ:hartree-energy-dens}, one obtains
\begin{equation}
    \epsilon_\mathrm{H}(k_F, q_F) = \frac{\sigma}{2\mu} \left\{ \frac{k_F^3}{2\pi^2} - I(k_F, q_F) \right\}\label{equ:potenergydens}
\end{equation}

where the integral $I(k_F, q_F)$ gives:
\begin{align}
    I(k_F, q_F) &= \frac{\mu}{4\pi^3} \left\{ \frac{2}{3} q_F k_F + \left( \frac{k_F^2}{2} + \frac{\mu^2}{3}\right) \ln{\left[ \frac{(q_F-k_F)^2 + \mu^2}{(q_F + k_F)^2 + \mu^2} \right]} \color{white}\right\}\\
    &+ \color{white}\left\{ \color{black}\left( \frac{2}{3} \frac{q_F^3}{\mu} - 6\mu q_F \right) \arctan{\left[ \frac{2 k_F\mu}{\mu^2 + q_F^2 - k_F^2} \right]} + \frac{2}{3} \frac{k_F^3}{\mu} \arctan{\left[ \frac{2 q_F\mu}{\mu^2 + k_F^2 - q_F^2} \right]} \right\}
\end{align}

This expression breaks the symmetry with respect to the Fermi momenta $q_F$ and $k_F$, even though $W(\bm{k}- \bm{q})$ had the symmetric form. 
In a symmetric system of light quarks only the Fermi momenta of the interacting quark species are equal ($k_F=q_F)$ so that the above expression further simplifies to
\begin{align}
    I(k_F, k_F) &= \frac{\mu}{4\pi^3} \left\{ \frac{2}{3} k_F^2 + \left( \frac{k_F^2}{2} + \frac{\mu^2}{3}\right) \ln{\left[ \frac{\mu^2}{4k_F^2 + \mu^2} \right]} + \left( \frac{4}{3} \frac{k_F^3}{\mu} - 6\mu k_F \right) \arctan{\left[ \frac{2 k_F\mu}{\mu^2} \right]} \right\} \label{equ:momentumeq}
\end{align}

In the limit $\mu\to \infty$, this integral vanishes and thus the non-interacting case is established, since the Hartree energy density functional 
\eqref{equ:potenergydens} vanishes.  
In the opposite limit $\mu\to 0$, without any screening, this result diverges (as to be expected) and can therefore not readily be applied in calculations of the equation of state of quark matter with confinement interactions.
We note that the density functional obtained in Ref. \cite{Shukla:2025egs} differs from our result since a different form for the confinement potential in momentum space has been assumed and an expansion has been applied.
%\color{red}I would say, they applied an approximation.
%\color{black}

\subsection{Towards a potential energy density functional}
The main reason for the divergence of the density functional in the case of unscreened confinement is that the assumption that all quarks interact with each other via the free-space confining potential is not justified. Since the interaction potential rises linearly with the distance of the interacting quarks, there would be infinite energy in those strings drawn between the most distant quarks, when the volume goes to infinity in the thermodynamic limit.
In order to account for the confining property of the interaction between color charges (quarks and gluons), this interaction shall get saturated once it forms a color neutral correlation. This principle has also been formulated as the absence of color van der Waals forces.
In order to minimize the total free energy of the many-quark system (for which the many-particle concept of a density functional shall be applied), the total length of the strings drawn between color neutral pairs (in mesonic correlations) or triples (in baryonic correlations) shall be minimal, see Ref.~\cite{Ropke:1986qs} for further details.  
A principle which would fulfill this constraint is the saturation of color interactions within the nearest-neighbor distance. It is realized in the string-flip model either by numerical simulation \cite{Horowitz:1985tx} or by adopting a statistical distribution of nearest-neighbor distances \cite{Ropke:1986qs}.

In the present work, where we have already introduced with the regularization parameter $\mu$ an effective Debye screening radius 
$r_D=1/\mu$, we can employ this parameter to obtain a finite result for the density functional. To this end, we apply the concept of a Debye sphere and demand that at the surface of this sphere around a quark of given color another quark of matching anti-color can be found. 
This is important, as in the present work we assume a color neutral and spin neutral system. Restricting the system in such a way reduces number of configurations and this in turn manifests itself in a reduction of the configuration space $\mathbb Q$. Due to the fact that only one particle can sit on the Debye sphere's surface, the configuration space is three dimensional. 
We might wonder about the probability density $n(\mathbb )$ that another potential partner sits at a distance $r_D$ relative to the sphere's center. The latter should be normalized to unity if one takes the configuration space integral:
\begin{equation}
    \frac{1}{2N_c} \int_{\mathbb R^3} d\mathbb Q\; n(\mathbb Q) = \frac{n}{2N_c} \int_{0\leq |\bm{r}| \leq r_D} d^3r\; = \frac{n}{2N_c} \frac{4}{3} \pi r_D^3 = 1
\end{equation}
As a consequence the probability density of finding another particle in distance $r_D$ is given by
\begin{equation}
    n = \frac{3N_c}{2\pi}\frac{1}{r_D^3}
\end{equation}

For applications at $T=0$, we use the relation of quark density with the quark Fermi momentum
\begin{equation}
\label{eq:density}
    n = 2N_f N_c \int \frac{d^3\bm{k}}{(2\pi)^3} n_F(\bm{k}) 
    = \frac{2 N_f N_c}{6\pi^2} k_F^3~,
\end{equation}
to obtain the relation between screening parameter and Fermi momentum:
\begin{equation}
\label{eq:mu}
    \mu = \frac{1}{r_D}
    = \left( \frac{3 N_c}{2 \pi} \right)^{-\frac{1}{3}} \left( \frac{2 N_f N_c}{6\pi^2} k_F^3 \right)^{\frac{1}{3}}
    = \xi k_F
\end{equation}
For simplicity, we have introduced $\xi$ which is given as follows:
\begin{equation}
    \xi = \left(\frac{2 N_f}{9\pi}\right)^{1/3}
\end{equation}
We may now insert $\mu = \xi k_F$ into equation \eqref{equ:potenergydens} with \label{equ:momentumeqref} which leads to:
\begin{equation}
    U(k_F, k_F) = \frac{\sigma \chi}{2} k_F^2
\end{equation}
with
\begin{equation}
    \chi = \frac{1}{2\pi^2 \xi} - \frac{1}{4\pi^3} \left\{ \frac{2}{3} + \left( \frac{1}{2} + \frac{\xi^2}{3} \right) \ln{\left[ \frac{\xi^2}{4 + \xi^2} \right]} 
    + \left( \frac{4}{3\xi} - 6\xi\right) \arctan{\left[ \frac{2}{\xi} \right]}
    \right\}
\end{equation}

\begin{figure}[H]
    \centering
    \includegraphics[width=\linewidth]{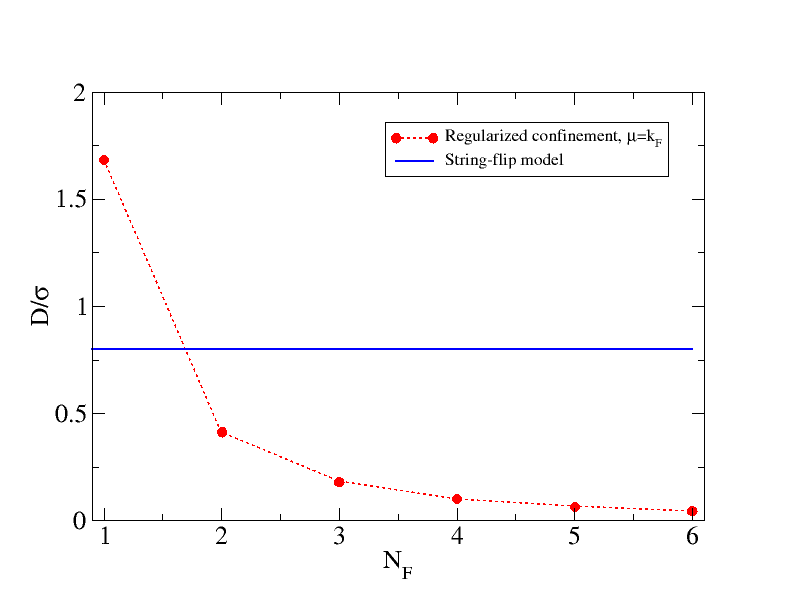}
    \caption{Screening factor $D/\sigma$ in the confining interaction energy density functional as a function of the number of quark flavors $N_f$. For comparison, the corresponding value for the string-flip model \cite{Barter:1992rg} is shown.}
    \label{fig:d_over_flav}
\end{figure}

Because the relation between density and Fermi momentum in equation \eqref{eq:density} still holds, we may view the potential energy density $U$ as a functional of the density $n$. With this we obtain our result for the density functional of the interaction energy density of quark matter with saturated confinement interactions
\begin{align}
    U[n] &= \left( \frac{6 \pi^2}{2N_c N_f} \right)^{\frac{2}{3}} \frac{\sigma \chi}{2}  n^{\frac{2}{3}}\\
    &=  D\, n^{2/3}\label{equ:df-u}
\end{align}
where the constant 
\begin{equation}
    D= \left( \frac{6 \pi^2}{2N_c N_f} \right)^{\frac{2}{3}} \frac{\chi}{2} \sigma
\end{equation}
is an effective string tension. Typical values of the coefficients $\xi$, $\chi$ and $D$ for $N_c = 3$ are given in Table \ref{tab:coeff} depending on the number of flavors. Figure \ref{fig:d_over_flav} shows this dependency and compares it with the value obtained for the string-flip model 
\cite{Barter:1992rg}.

\begin{table}[H]
\caption{Parameter values for different numbers of flavors for a system with three colors.}
    \centering
    \begin{tabular}{c|c|c|c}
\hline
$N_f$ & $\xi$ & $\chi$ & $D/\sigma$\\\hline
1 & 2.4180 & 0.1036 & 1.6825 \\
2 & 1.9192 & 0.1016 & 0.4123 \\
3 & 1.6765 & 0.1003 & 0.1810 \\
4 & 1.5232 & 0.0995 & 0.1009 \\
5 & 1.4140 & 0.0989 & 0.0642 \\
6 & 1.3307 & 0.0985 & 0.0444 \\ \hline
\end{tabular}
\label{tab:coeff}
\end{table}

Note that the effective string tension reduces as the number of quark flavors is increased.

In practical applications, one may want to account for a "melting" of the string tension in the dense medium and apply some reduction of the parameter $\sigma$ as, e.g., in \cite{Kaltenborn:2017hus}.

\section{Summary and Outlook}

We have performed the derivation of a density functional for quark matter with linear confinement interaction, applying
the principle of saturation of color interactions. For the saturation parameter $\mu$ the ansatz $\mu=k_F$ is employed
which leads to an interaction energy density functional $\propto n^{2/3}$ that has been introduced before on purely phenomenological grounds using a nearest-neighbor distance argument. 
In a future work, we would like to elaborate on the string breaking argument for the saturation of the confinement potential which would suggest that $\sigma/\mu=2 m$ should be a reasonable ansatz, whereby $m$ should be the dynamically generates quark mass as a solution of the mass gap equation for the saturated confinement potential. 

\subsection*{Acknowledgement}
We acknowledge discussions with Oleksii Ivanytskyi, Biplab Mahato and Zhandos Moldabekov on different aspects of the regularized confinement potential.

%\end{document}

%\bibliography{conf}

\begin{appendix}

\section{Fourier transform of the screened confinement potential}
\label{app:fourier}
The derivation of the Fourier transform of the linear confinement potential makes use of the fact that it can be obtained from a screened potential \eqref{eq:cornell-sc} in the limit of a vanishing screening parameter $\mu\to 0$ (infinite Debye radius) \cite{Dixit:1989vq}.

Considering the regularization
\begin{equation}
    V_1(\bm{r}) = \lim_{\mu \to 0} \frac{\sigma}{\mu} \left( 1 - e^{-\mu r} \right)
\end{equation}
whose Fourier transform
\begin{equation}
    V_1(\bm{k}) = \int d^3\bm{r}\; V_1(\bm{r}) e^{i \bm{k}\cdot\bm{r}} = \lim_{\mu\to 0} \frac{\sigma}{\mu} \int d^3\bm{r}\; \left[ 1 - e^{-\mu r} \right] e^{i \bm{k}\cdot\bm{r}}
\end{equation}
can be calculated using the Gamma function.
Firstly, identify the Dirac delta function
\begin{equation}
    (2\pi)^3\delta^{(3)}(\bm{k}) = \int d^3\bm{r}\; e^{i \bm{k}\cdot\bm{r}} 
\end{equation}
and rewrite the Fourier transform into
\begin{equation}
    V_1(\bm{k}) = \lim_{\mu\to 0} \frac{\sigma}{\mu} \left[(2\pi)^3\delta(\bm{k})^{(3)} - I(\bm{k})\right]
\end{equation}
where
\begin{equation}
    I(\bm{k}) = \int d^3\bm{r}\; e^{-\mu r} e^{i \bm{k}\cdot\bm{r}}
\end{equation}
Then let $r:=|\bm{r}|$ and $k := |\bm{k}|$ and transform $I(\bm{k})$ to spherical coordinates. Rewriting of the dot product results in
\begin{equation}
    I(\bm{k}) = \int_0^{2\pi} d\varphi \int_0^\pi d\theta \int_0^\infty d r\; r^2 \sin{\theta}\, e^{-\mu r} e^{i k r \cos{\theta}}
\end{equation}
where the integral over $\varphi$ simply yields a factor of $2\pi$. By substituting $z=\cos{\theta}$, the integral can be cast into
\begin{equation}
    I(\bm{k}) = 2\pi \int_0^\infty d r\; r^2 e^{-\mu r} \int_{-1}^1 d z\; e^{i k r z}
\end{equation}
where the integral over $z$ evaluates to the following:
\begin{equation}
    \int_{-1}^1 d z\; e^{i k r z} = \frac{2}{kr} \sin{(kr)}\label{equ:intoverz}
\end{equation}
After simplification and the identification of $\sin{(kr)}$ with $\mathrm{Im}\left[e^{ikr}\right]$, one is left with the expression
\begin{equation}
    I(\bm{k}) = \frac{4\pi}{k} \mathrm{Im}\left[\int_0^\infty dr\; r e^{-(\mu - ik) r}\right]
\end{equation}
From there, the substitution $x = (\mu - ik)r$ turns the integrand into a form which allows the identification with the Gamma function:
\begin{equation}
    I(\bm{k}) = \frac{4\pi}{k} \mathrm{Im}\left[ \frac{1}{(\mu - ik)^2}\int_0^\infty dx\; x e^{-x}\right] = \frac{4\pi}{k} \mathrm{Im}\left[ \frac{1}{(\mu - ik)^2}\Gamma(2)\right]
\end{equation}
With this argument the Gamma function evaluates to $1$, hence taking the imaginary part leads to the simple result of:
\begin{equation}
    I(\bm{k}) = \frac{8\pi \mu}{(\mu^2 + k^2)^2}
\end{equation}
Piecing everything together, the 1D regularized confinement potential's Fourier transform is given by the following expression:
\begin{equation}
    V_1(\bm{k}) = \lim_{\mu\to 0} \frac{\sigma}{\mu} \left[(2\pi)^3\delta(\bm{k})^{(3)} - \frac{8\pi \mu}{(\mu^2 + k^2)^2}\right]
\end{equation}

\section{Sixfold momentum space integration}\label{app:trafo_sub}

The six dimensional integral of the form
\begin{align}
    U_1(k_F,q_F)&= \frac{1}{2}\int \frac{d^3\bm{k}}{(2\pi)^3}
    \int \frac{d^3\bm{q}}{(2\pi)^3} V_1(\bm{k}-\bm{q}) n(\bm{k})n(\bm{q}) \\
    &= \frac{1}{2} \int_{|\bm{k}|\le k_F} \frac{d^3k}{(2\pi)^3}
    \int_{|\bm{q}|\le q_F} \frac{d^3q}{(2\pi)^3} V_1(\bm{k}-\bm{q})
\end{align}
may be tackled by firstly integrating the Dirac delta function over three momenta $q$ and $k$ which yields:
\begin{equation}
    (2\pi)^3 \int_{|\bm{k}|\le k_F} \frac{d^3k}{(2\pi)^3}
    \int_{|\bm{q}|\le q_F} \frac{d^3q}{(2\pi)^3} \delta^{(3)}(\bm{k} - \bm{q}) = \int_{|\bm{k}|\le k_F} \frac{d^3k}{(2\pi)^3} = \frac{k_F^3}{2\pi^2} 
\end{equation}
Secondly, it is best to focus on integrating the second term over $q$. For this let
\begin{equation}
    I(k, q_F) := \int \frac{d^3\bm{q}}{(2\pi)^3} \frac{8\pi \mu}{(\mu^2 + (\bm{k} - \bm{q})^2)^2} = \int \frac{d^3\bm{q}}{(2\pi)^3} \frac{8\pi \mu}{(\mu^2 + (k^2 + q^2 - 2kq \cos{\theta})^2}
\end{equation}
and transform to spherical coordinates:
\begin{equation}
    I(k, q_F) = \frac{\mu}{\pi^2} \int_0^{2\pi} d\varphi \int_0^\pi d\theta \int_0^{q_F} dq\; q^2 \sin{\theta} \frac{1}{(\mu^2 + k^2 + q^2 - 2kq \cos{\theta})^2}
\end{equation}
The integration over $\varphi$ simply contributes a factor of $2\pi$. Substituting $z=\cos{\theta}$ and factoring out $(2kq)^{-2}$ turns $I(k, q_F)$ into
\begin{equation}
    I(k, q_F) = \frac{\mu}{2\pi}\frac{1}{k^2} \int_0^{q_F} dq \int_{-1}^{1} dz\; \frac{1}{(z - z_0(k, q))^2} \quad \mathrm{where} \quad z_0(k, q) = \frac{\mu^2 + k^2 + q^2}{2kq}\label{equ:1d_z_begin}
\end{equation}
Subsequent substitution of $u = z-z_0(k, q)$ massages $I(k, q_F)$ into the following form
\begin{equation}
    I(k, q_F) = \frac{\mu}{2\pi}\frac{1}{k^2} \int_0^{q_F} dq \int_{-1 - z_0(k, q)}^{1 - z_0(k, q)} du\; \frac{1}{u^2}
\end{equation}
where the integral over $u$ yields:
\begin{align}
    \int_{-1 - z_0(k, q)}^{1 - z_0(k, q)} du\; \frac{1}{u^2} &= \frac{1}{z_0-1} - \frac{1}{z_0 + 1}\\
    &= \frac{2kq}{\mu^2 + (k-q)^2} - \frac{2kq}{\mu^2 + (k+q)^2}\label{equ:uint}
\end{align}
To finish the integration over three momentum $\bm{q}$, one has to evaluate
\begin{equation}
    I_\mp(k, q_F) := \int_0^{q_F} dq\; \frac{2kq}{\mu^2 + (k\mp q)^2}
\end{equation}
For this, attention should be paid to the numerator's and the denominator's powers in $q$. Let $f(q) := \mu^2 + (k\mp q)^2$, then it is possible to express the numerator as $2kq = k f'(q) \pm 2k^2$. This enables $I_\mp$ to be rewritten as follows
\begin{align}
    I_\mp (k, q_F) &= k \int_0^{q_F} dq\; \frac{f'(q)}{f(q)} \pm 2k^2 \int_0^{q_F} \frac{dq}{\mu^2 + (k \mp q)^2}\\
    &= k \ln{\left[ \frac{\mu^2 + (k \mp q_F)^2}{\mu^2 + k^2} \right]} - \frac{2k^2}{\mu} \arctan{\left( \frac{k\mp q_F}{\mu} \right)} \pm \frac{2k^2}{\mu} \arctan{\left( \frac{k}{\mu} \right)}
\end{align}
and thus $I(k, q_F)$ evaluates to
\begin{equation}
    I(k, q_F) = \frac{\mu}{2\pi} \left\{ \frac{1}{k} \ln{\left[ \frac{\mu^2 + (k^2-q_F)^2}{\mu^2 + (k+q_F)^2} \right]} + \frac{2}{\mu} \arctan{\left[ \frac{2 q_F \mu}{\mu^2 + k^2 - q_F^2} \right]} \right\}
\end{equation}
where use was made of
\begin{equation}
    \arctan{(a)} - \arctan{(b)} = \arctan{\left(\frac{a-b}{1+ab}\right)}
\end{equation}
Integration over three momenta $\bm{k}$ is a matter of evaluating
\begin{equation}
    I(k_F, q_F) = \int_{|\bm{k}|\le k_F} \frac{d^3\bm{k}}{(2\pi)^3} \;I(k, q_F) = \frac{1}{2\pi^2} \int_0^{k_F} dk\; k^2 I(k, q_F)
\end{equation}
For this one has to calculate
\begin{equation}
    I_1(k_F, q_F) = \int_0^{k_F} dk\; k \ln{\left[ \frac{\mu^2 + (k^2-q_F)^2}{\mu^2 + (k+q_F)^2} \right]}
\end{equation}
as well as
\begin{equation}
    I_2(k_F, q_F) = \frac{2}{\mu}\int_0^{k_F} dk\; k^2 \arctan{\left[ \frac{2 q_F \mu}{\mu^2 + k^2 - q_F^2} \right]}
\end{equation}
In both cases, the interplay between integration by parts, polynomial division and contour integration is the way to go.
Focusing on $I_1(k_F, q_F)$, it is possible to write
\begin{align}
    I_1(k_F, q_F) &= \left.\frac{k^2}{2}  \ln{\left[ \frac{\mu^2 + (k^2-q_F)^2}{\mu^2 + (k+q_F)^2} \right]}\right\vert_0^{k_F}\\
    &- \int_0^{k_F} dk\; k^2 \frac{k-q_F}{\mu^2 + (k-q_F)^2} + \int_0^{k_F} dk\; k^2 \frac{k+q_F}{\mu^2 + (k+q_F)^2}
\end{align}
and by a subsequent polynomial division, the two integrals can be cast into the following:
\begin{align}
    \int_0^{k_F} dk\; k^2 \frac{k-q_F}{\mu^2 + (k-q_F)^2} &= \frac{k_F^2}{2} + q_F k_F + \int_0^{k_F} dk\; \frac{(q_F^2 - \mu^2)k - (\mu^2q_F + q_F^3)}{\mu^2 + (k-q_F)^2}\\
    \int_0^{k_F} dk\; k^2 \frac{k+q_F}{\mu^2 + (k+q_F)^2} &=\frac{k_F^2}{2} -q_F k_F + \int_0^{k_F} dk\; \frac{(q_F^2 - \mu^2)k + (\mu^2q_F + q_F^3)}{\mu^2 + (k+q_F)^2}
\end{align}
Evaluating integrals of the form
\begin{align}
    I_{\mp} &:= \int_0^{k_F} dk\; \frac{(q_F^2 - \mu^2)k \mp (\mu^2q_F + q_F^3)}{\mu^2 + (k\mp q_F)^2}\\
    &= (q_F^2 - \mu^2) \int_0^{k_F} dk\; \frac{k}{\mu^2 + (k\mp q_F)^2} \mp (\mu^2 q_F + q_F^3) \int_0^{k_F} \frac{dk}{\mu^2 + (k\mp q_F)^2} 
\end{align}
is simply a matter of substituting $f_\mp(k) = \mu^2 + (k \mp q_F)^2$ and combining terms into
\begin{equation}
    I_{\mp} = \frac{(q_F^2 - \mu^2)}{2} \int_0^{k_F} dk\; \frac{f'_\mp(k)}{f_\mp(k)} \mp 2\mu^2 q_F \int_0^{k_F} \frac{dk}{\mu^2 + (k\mp q_F)^2} 
\end{equation}
Recognizing the closed form for the natural logarithm and the second integrand's similarity to the derivative of $\mathrm{arctan}$ one obtains
\begin{equation}
    I_{\mp} = \frac{(q_F^2 - \mu^2)}{2} \ln{\left[ \frac{\mu^2 + (k_F \mp q_F)^2}{\mu^2 + q_F^2} \right]} \mp 2\mu q_F \left[ \mathrm{arctan}\left( \frac{k_F \mp q_F}{\mu} \right) \pm \mathrm{arctan}\left( \frac{q_F}{\mu} \right) \right]
\end{equation}
With this out of the way, $I_1(k_F, q_F)$ evaluates to the following mess:
\begin{equation}
    I_1(k_F, q_F) = \frac{k_F^2 - q_F^2 + \mu^2}{2} \ln{\left[ \frac{\mu^2 + (k_F - q_F)^2}{\mu^2 + (k_F + q_F)^2} \right]} + 2\mu q_F\,\mathrm{arctan}\left[ \frac{2 k_F \mu}{\mu^2 - k_F^2 + q_F^2} \right] - 2q_F k_F
\end{equation}
Focusing on $I_2(k_F, q_F)$, one use integration by parts to write:
\begin{equation}
    I_2(k_F, q_F) = \frac{2 k_F^3}{3\mu} \frac{k_F^3}{3} \mathrm{arctan} \left( \frac{2 q_F \mu}{ \mu^2 + k_F^2 - q_F^2} \right) + \frac{8}{3} q_F \underbrace{\int_0^{k_F} dk\; \frac{k^4}{[\mu^2 + (k-q_F)^2][\mu^2 + (k+q_F)^2]}}_{:=W(k_F, q_F)}
\end{equation}
Now the integral $W(k_F, q_F)$ is the object of desire. By performing polynomial long division it may be massaged into
\begin{align}
    W(k_F, q_F) = k_F - 2(\mu^2 - q_F^2) \underbrace{\int_0^{k_F} dk\; g_1(k_, q_F)}_{:=W_1(k_F, q_F)} - (\mu^2 + q_F^2)^2 \underbrace{\int_0^{k_F}dk\; g_2(k_, q_F)}_{:=W_2(k_F, q_F)}
\end{align}
where the integrands are given by:
\begin{align}
    g_1(k, q_F) &= \frac{k^2}{[\mu^2 + (k-q_F)^2][\mu^2 + (k+q_F)^2]}\\
    g_2(k, q_F) &= \frac{1}{[\mu^2 + (k-q_F)^2][\mu^2 + (k+q_F)^2]}
\end{align}
The integrals $W_1(k_F, q_F)$ and $W_2(k_F, q_F)$ may now be evaluated using contour integration.\\
For this notice how any definite integral $\int_0^a dx\; f(x)$ may be written as an improper integral if one weighs the integrand with a regulating distribution. To be more specific, this regulating distribution is a distribution that activates $f(x)$ in the desired interval of integration and deactivates it everywhere else. Cauchy's residue theorem requires the integrand to be analytic, hence our regulating distribution cannot be a product of Heavyside step functions. Instead one may choose the difference of two Fermi-Dirac distributions
\begin{equation}
    f(x-a) - f(x) =\frac{1}{1+ e^{\beta (x - a)}} - \frac{1}{1+ e^{\beta x}} := \omega_\beta(x, a)
\end{equation}
as a regulating function and take the limit of $\beta\to\infty$ after applying the residue theorem. This theorem demands knowledge of $g_1(k_F, q_F)$'s and $g_2(k_F, q_F)$'s poles. In both cases they lie at
\begin{align}
    k_1 &= q_F + i\mu \qquad k_2 = \bar{k}_1 = q_F - i\mu\\
    k_3 &= -q_F + i\mu \qquad k_4 = \bar{k}_3 = -q_F - i\mu
\end{align}
Hence 
\begin{align}
    W_i(k_F, q_F) = \lim_{\beta\to\infty} \int_{-\infty}^{\infty} dk\; g_i(k, q_F) \omega_\beta (k, k_F)
\end{align}
Due to the regulating distribution being a difference of two Fermi-Dirac distributions, $W_1(k_F, q_F)$ and $W_2(k_F, q_F)$ split into a differences of integrals as governed by $\omega_\beta(k, k_F)$. At this point we analytically continue $\omega_\beta(k, k_F)$ to $\omega_\beta(z, k_F)$. As a result, the latter has poles at the Matsubara frequencies $\nu_n = (2n+1)i\pi/\beta + k_F$ and $\mu_n = (2n+1)i\pi/\beta$. 

\begin{figure}[h!] 
  \centering
  \begin{subfigure}[t]{0.6\textwidth}
    \centering
    \includegraphics[width=\linewidth]{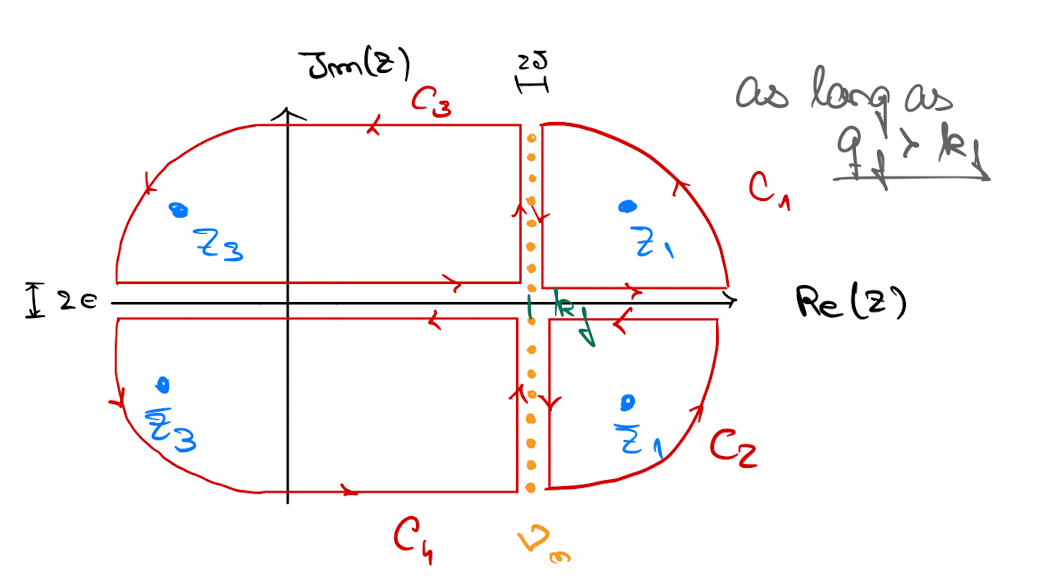}
    \caption{$g_i(z)f(z-k_F)$}
    \label{fig:sub2}
  \end{subfigure}\hfill
    \begin{subfigure}[t]{0.38\textwidth}
    \centering
    \includegraphics[width=\linewidth]{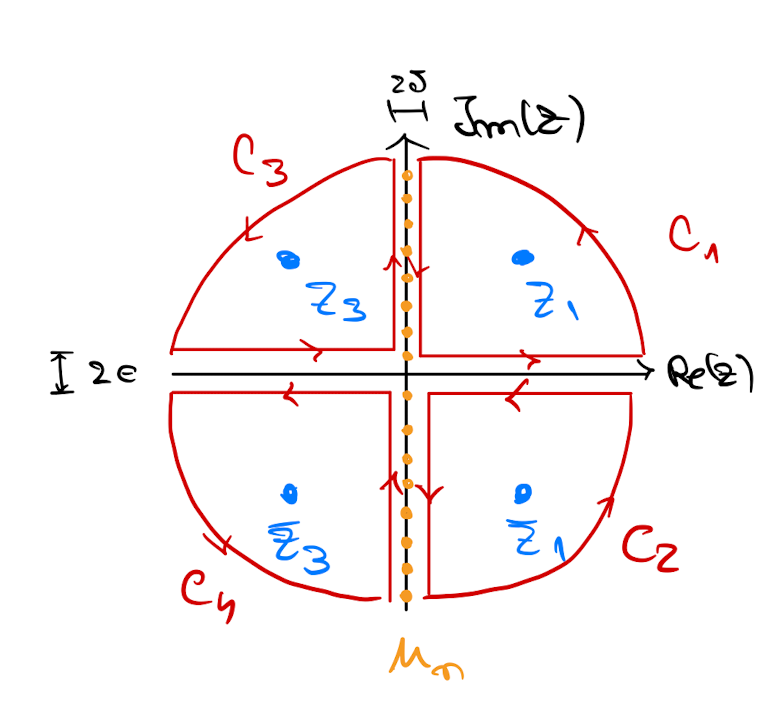}
    \caption{$g_i(z)f(z)$}
    \label{fig:sub1}
  \end{subfigure}
  \caption{Complex planes of the integrands and the chosen contours.}
  \label{fig:both}
\end{figure}

Figure \ref{fig:both} depicts the complex planes of the functions $g_i(z)f(z-k_F)$ and $g_i(z)f(z)$.
In each case, we choose the four contours $C_1$, $C_2$, $C_3$ and $C_4$ that encircle each of the poles $z_1$, $z_2$, $z_3$ and $z_4$ respectively.

\begin{figure}
    \centering
    \includegraphics[width=0.7\linewidth]{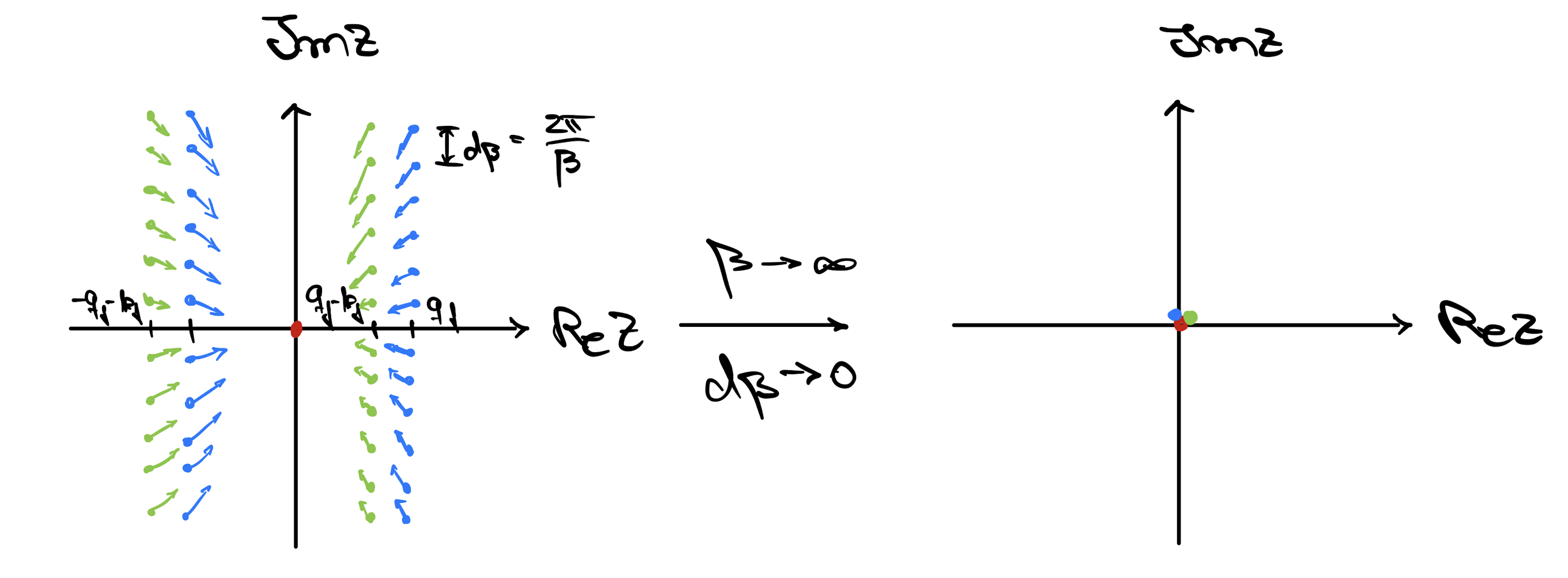}
    \caption{Poles of $u_\beta(z_1, \bar{z}_3)$ flowing towards the origin of the complex plane as $\beta\to\infty$.}
    \label{fig:flow}
\end{figure}

As we integrate on the interval $(-\infty, \infty)$, it is completely unnecessary to parametrize the arcs as their contribution will vanish according to Jordan's lemma \cite{Arfken2013}. In the limit of $(\epsilon, \delta)\to(0, 0)$ one may write the following:
\begin{align}
    \int_{-\infty}^\infty dk\; g_i(k) \omega_\beta(k, k_F) = \sum_{j=1}^4 \oint_{C_j} dz\; g_i(z) \omega_\beta(z, k_F) = 2\pi i \sum_{j=1}^4 \mathrm{Res}_{z_j} [g_i(z)\omega_\beta(z, k_F)]
\end{align}
As the task is now to calculate the residues at simple poles using
\begin{equation}
    \mathrm{Res}_{z_j} [g_i(z)\omega_\beta(z, k_F)] = \lim_{z\to z_j} [(z-z_j) g_i(z) \omega_\beta(z, k_F)]
\end{equation}
in the limit of $\beta\to\infty$ one obtains the following very similar expressions:
\begin{align}
    W_1(k_F, q_F)= \frac{\pi}{8\mu q_F} \lim_{\beta\to\infty} [&z_1 u_\beta(z_1, \bar{z}_3, k_F) + \bar{z}_1 u_\beta(\bar{z}_1, z_3, k_F) ]\\
    W_2(k_F, q_F)= \frac{\pi}{8\mu q_F(q_F^2 + \mu^2)} \lim_{\beta\to\infty} [&z_1 u_\beta(\bar{z}_1, z_3, k_F) + \bar{z}_1 u_\beta(z_1, \bar{z}_3, k_F)]
\end{align}
For convenience we have defined the function
\begin{equation}
    u_\beta(a, b, c) := \omega_\beta(a, c) + \omega_\beta(b, c)
\end{equation}
Focusing on $u_\beta(z_1, \bar{z}_3, k_F)$, we may guess its $\beta\to\infty$ limit by carefully observing what happens to the pole structure. We also have to keep in mind its co-domain which we know is given by $0\leq u_\beta(z_1, \bar{z}_3, k_F)\leq 2$. The upper bound of $2$ is explained by adding two functions $\omega_\beta(\cdot, c)$ which are separately upper bounded by $1$. The distribution $\omega_\beta(z_1, k_F)$ has poles at
\begin{equation}
    \frac{(2n+1)\pi}{\beta} = q_F - k_F + i\mu \quad \text{and} \quad \frac{(2n+1)\pi}{\beta} = q_F + i\mu
\end{equation}
whereas $\omega_\beta(\bar{z}_3, k_F)$ has poles at 
\begin{equation}
    \frac{(2n+1)\pi}{\beta} = -q_F - k_F - i\mu \quad \text{and} \quad \frac{(2n+1)\pi}{\beta} = -q_F - i\mu
\end{equation}
As $\beta\to\infty$ all poles flow towards the origin of the complex plane at $z=0$. This is depicted in Figure \ref{fig:flow}.
The only function with a branch point at $z=0$ is $\ln(z)$ where $\mathrm{Re}(z)\geq0$. From this we can safely say that the $\beta\to\infty$ limit of $u_\beta(z_1, \bar{z}_3, k_F)$ is given by
\begin{align}
    \lim_{\beta\to\infty} u_\beta(z_1, \bar{z}_3, k_F) &= \frac{1}{\pi i} [\ln{(q_F - k_F + i\mu)} - \ln{(q_F + k_F + i\mu)} - \ln{(q_F + i\mu)} + \ln{(q_F + i\mu)}]\\
    &= \frac{1}{\pi i} \ln{\left[ \frac{q_F - k_F + i\mu}{q_F + k_F + i\mu} \right]} = \frac{2}{\pi} \mathrm{arctan}\left[\frac{k_F}{\mu - iq_F}\right]
\end{align}
whose co-domain obeys the restriction from above.
The $\beta\to\infty$ limit of $u_\beta(\bar{z}_1, z_3, k_F)$ can be obtained by using the fact that $u_\beta(\bar{z}_1, z_3, k_F)$ and $u_\beta(z_1, \bar{z}_3, k_F)$ are connected via complex conjugation, namely: $u_\beta(\bar{z}_1, z_3, k_F) = \bar{u}_\beta (z_1, \bar{z}_3, k_F)$. This means that
\begin{align}
    \lim_{\beta\to\infty} u_\beta(\bar{z}_1, z_3, k_F) &= \lim_{\beta\to\infty} \bar{u}_\beta(z_1, \bar{z}_3, k_F)\\
    &= -\frac{1}{\pi i} \ln{\left[ \frac{q_F - k_F - i\mu}{q_F + k_F - i\mu} \right]} = \frac{2}{\pi} \mathrm{arctan}\left[\frac{k_F}{\mu + iq_F}\right]
\end{align}
Very beautiful.
With the $\beta\to\infty$ limits known, we can puzzle together the magnificent closed forms of $W_1(k_F, q_F)$ and $W_2(k_F, q_F)$:
\begin{align}
    W_1(k_F, q_F)= \frac{1}{4\mu q_F} \left\{ q_F \mathrm{arctan}\left[ \frac{2\mu k_F}{q_F^2 - k_F^2 + \mu^2} \right] + \frac{\mu}{2} \ln{\left[ \frac{\mu^2 + (q_F - k_F)^2}{\mu^2 + (q_F + k_F)^2} \right]} \right\}\\
    W_2(k_F, q_F)= \frac{1}{4\mu q_F}\frac{1}{(\mu^2 + q_F^2)} \left\{ q_F \mathrm{arctan}\left[ \frac{2\mu k_F}{q_F^2 - k_F^2 + \mu^2} \right] - \frac{\mu}{2} \ln{\left[ \frac{\mu^2 + (q_F - k_F)^2}{\mu^2 + (q_F + k_F)^2} \right]} \right\}
\end{align}
This means that
\begin{align}
    W(k_F, q_F) = k_F - \frac{3\mu^2 - q_F^2}{4\mu} \mathrm{arctan}\left[ \frac{2\mu k_F}{q_F^2 - k_F^2 + \mu^2} \right] - \frac{\mu^2 - 3 q_F^2}{8 q_F} \ln{\left[ \frac{\mu^2 + (q_F - k_F)^2}{\mu^2 + (q_F + k_F)^2} \right]}
\end{align}
and hence $I_2(k_F, q_F)$ evaluates to the following expression:
\begin{align}
    I_2(k_F, q_F) &= \frac{2k_F^3}{3\mu} \mathrm{arctan}\left( \frac{2q_F \mu}{\mu^2 + k_F^2 - q_F^2} \right)\\
    &+ \frac{8}{3}q_F \left\{ k_F + \frac{q_F^2 - 3\mu^2}{4\mu} \mathrm{arctan}\left[ \frac{2\mu k_F}{q_F^2 - k_F^2 + \mu^2} \right] + \frac{3 q_F^2 - \mu^2}{8 q_F} \ln{\left[ \frac{\mu^2 + (q_F - k_F)^2}{\mu^2 + (q_F + k_F)^2} \right]}  \right\}.
\end{align}

With $I_1(k_F, q_F)$ and $I_2(k_F, q_F)$ evaluated, it is finally time to piece everything together. In doing so, $I(k_F, q_F)$ evaluates to
\begin{align}
    I(k_F, q_F) &= \frac{\mu}{2\pi} \left\{ I_1(k_F, q_F) + I_2(k_F, q_F)\right\}\\
    &= \frac{\mu}{4\pi^3} \left\{ \frac{2}{3} q_F k_F + \left( \frac{k_F^2}{2} + \frac{\mu^2}{3}\right) \ln{\left[ \frac{(q_F-k_F)^2 + \mu^2}{(q_F + k_F)^2 + \mu^2} \right]} \color{white}\right\}\\
    &+ \color{white}\left\{ \color{black}\left( \frac{2}{3} \frac{q_F^3}{\mu} - 6\mu q_F \right) \arctan{\left[ \frac{2 k_F\mu}{\mu^2 + q_F^2 - k_F^2} \right]} + \frac{2}{3} \frac{k_F^3}{\mu} \arctan{\left[ \frac{2 q_F\mu}{\mu^2 + k_F^2 - q_F^2} \right]} \right\}
\end{align}
and hence
\begin{equation}
    U_1(k_F, q_F) = \frac{\sigma}{2\mu} \left\{ \frac{k_F^3}{2\pi^2} - I(k_F, q_F) \right\}.
\end{equation}

\section{Nearest neighbor distribution (string-flip model)}

The concept of the saturation of the string-type confinement interaction \eqref{eq:cornell-p} has been introduced in Ref. \cite{Blaschke:1984yj} (see also Refs. \cite{Lenz:1985jk,Horowitz:1985tx}) in order to remove divergencies that would occur if an unscreened confinement interaction would be used to evaluate the energy density of these interactions in a many-quark system. 
Here, we review this concept of the saturation of confinement within nearest-neighbors, following closely the notation of Ref. \cite{Barter:1992rg} in order to arrive at a density functional given as
\begin{equation}
\label{eq:df}
    U(n)=\frac{n^2}{N_c} \int V^{\rm eff}(r,n) d^3r,
\end{equation}
where the effective screened potential is given by
\begin{equation}
    V^{\rm eff}(r,n) = V(r) c(r,n)
\end{equation}
with the non-normalized nearest-neighbor distribution function in three dimensions for uncorrelated quarks of density $n$ is 
\begin{equation}
    c(r,n)=\exp\left(-\frac{4\pi\, n}{3N_c} r^3\right).
\end{equation}
The integral in the density functional \eqref{eq:df} can be analytically evaluated for interaction potentials of polynomial form and results for the linear confinement potential in
\begin{equation}
    U(n)=\sigma\, \Gamma\left(\frac{4}{3}\right) \left(\frac{3 N_c}{4\pi}\right)^{1/3}\, n^{2/3}~.
\end{equation}
This result has the same form as Eq. \eqref{equ:df-u}, but with $\Gamma({4}/{3})=0.89298$ the prefactor in the case of the string-flip model,
$\chi_{\rm SFM}=0.79895$, is much larger than the values of the order $10^{-2}$ in Table \ref{tab:coeff}.

\end{appendix}

\end{document}